\documentclass[aps,twocolumn,nofootinbib,preprintnumbers,superscriptaddress]{revtex4-1}
\pdfoutput=1
\usepackage{color}
\usepackage[usenames,dvipsnames,table]{xcolor}
\usepackage{graphicx,amsmath,amssymb,amsthm,multirow,array,bm,}
\usepackage[mathscr]{eucal}
\usepackage[bbgreekl]{mathbbol}
\usepackage{amsfonts}
\usepackage{hyperref}
\usepackage{tikz}
\usepackage{bbm}
\usepackage{ulem}
\usepackage{lipsum}

\hypersetup{
    pdfstartview={FitH},    
    pdftitle={Universal features of left-right entanglement entropy},    
    pdfauthor={Diptarka Das, Shouvik Datta},     
    colorlinks=true,       
    linkcolor=blue,          
    citecolor=red,        
    filecolor=magenta,      
    urlcolor=blue           
}
\definecolor{rust}{rgb}{0.8,0.2,0.2}
\def\DD#1{{\color{black}{#1}}}

\def\jm#1{{\color{black}{#1}}}

\def\ket#1{\mid \! #1\rangle}
\def\bra#1{\langle \, #1 \! \mid\! \ }

\def\nn{\nonumber}
\def\pd{\partial}

\def\l1{{\text{1-loop}}}

\def\n1{\Bigg|_{n=1}}

\def\n{{(n)}}
\def\tr{\text{Tr}}

\def\bra#1{{\langle}#1|}
\def\ket#1{|#1\rangle}

\def\kket#1{|#1\rangle\hspace*{-.08cm}\rangle}

\def\cM{\mathcal{M}}
\def\cH{\mathcal{H}}
\def\cD{\mathcal{D}}
\def\qdim{\mathbbm{d}}
\def\cZ{\mathcal{N}}
\def\tq{\tilde{q}}
\def\ep{\epsilon}
\def\cS{\mathcal{S}}
\def\cB{\mathcal{B}}
\def\cO{\mathcal{O}}
\def\cH{\mathcal{H}}
\def\sS{\mathscr{S}}
\def\topo{{\text{topo}}}
\def\area{{\text{area}}}

\def\editx#1{\textcolor{black}{#1}}

\begin{document}

\title
{Universal features of left-right entanglement entropy 
}

\author{Diptarka Das}
\email{diptarka@physics.ucsd.edu}
\affiliation{Department of Physics, University of California at San Diego, 
 La Jolla, CA 92093, USA.}

\author{Shouvik Datta}
\email{shouvik@cts.iisc.ernet.in}
\affiliation{\mbox{Centre for High Energy Physics, Indian Institute of Science, C.~V.~Raman Avenue, Bangalore 560012, India.}}

\begin{abstract}
We show the presence of universal features in the entanglement entropy of regularized boundary states for (1+1)-$d$   conformal field theories on a circle when the reduced density matrix is obtained by tracing over right/left-moving modes.  We derive a general formula for the left-right entanglement entropy in terms of the central charge and the modular $\cS$ matrix of the theory.  When the state is chosen to be an Ishibashi state, this measure of entanglement is shown to precisely reproduce the spatial entanglement entropy of a (2+1)-$d$ topological quantum field theory.   We explicitly evaluate the left-right entanglement entropies for  the Ising model, the tricritical Ising model and the $\widehat{su}(2)_k$ WZW model as examples. 
\end{abstract}


\maketitle

\section{Introduction}
\label{sec:intro}
The idea of entanglement entropy, which originated in the context of explaining entropy of black holes \cite{Bombelli:1986rw}, has now become an established quantity of interest in quantum information, many-body physics \cite{Osterloh} and the AdS/CFT correspondence \cite{Ryu:2006bv}. This quantity, in its very essence, captures the quantum entanglement of a subset ($\cH_A$) of the Hilbert space ($\cH$) with the rest of it. Entanglement entropy is defined as the von-Neumann entropy corresponding to the reduced density matrix $\rho_A$, $S_E = -\tr (\rho_A \log \rho_A)$. The reduced density matrix of a subspace of $\mathcal{H}$  is, in turn, obtained by tracing out the degrees of freedom living in its complement ($\cH_{A'}$). 

In 1+1 dimensions, the entanglement entropy corresponding to a subsystem $A$ of a single-interval $l$, is known to obey the  {universal} formula : $S_A=(c/3) \log (l/ \epsilon)$ \cite{Holzhey:1994we,Calabrese:2004eu,Calabrese:2009qy}. This formula relates to `real-space entanglement', in the sense that the Hilbert space is geometrically partitioned.
In this article, we consider a special class of excited states for CFTs on a circle.  {In particular, we consider conformally invariant boundary states, $\ket{B}$, which are generically linear combinations of Ishibashi states, $\kket{h_a}$ \cite{ishibashi}.} This class of states appear in the context of  D-branes in string theory \cite{Polchinski,gaberdiel-lectures2}, critical quantum quenches  \cite{Calabrese:2006rx, Calabrese:2007rg, Takayanagi:2010wp, Cardy:2014rqa}, quantum impurity problems \cite{affleck-ludwig} and the TQFT/CFT correspondence \cite{haldane-li,ludwig-qi,chandran,rex-prb, Cano,swingle}.  {These states are in a basis  which is  holomorphically factorizable into their left  and right moving sectors. }The entanglement entropy associated with such a factorization is termed as  \textit{left-right entanglement entropy} (LREE).  The state $\ket{B}$ is, however, non-normalizable. A way to regularize the norm is to perform a Euclidean time evolution by $e^{-\epsilon H}$ and work with the regularized state $\ket{\cB} = ({e^{-\epsilon H}}/{{(\cZ_{B}) }^{1/2}})\ket{B}$ instead \cite{footnote01}. 
The left-right entanglement entropy was studied for states of this type for free bosons in \cite{zayas} and in XXZ chains in \cite{rex-prl}. 
In this work, we develop the formalism to calculate LREE for the  states, $\ket{\cB}$, in arbitrary CFTs. We investigate universalities determined by the typical CFT data \jm{and determine the dependence of this quantity on the choice of $\ket{\cB}$}.

 We shall see that for all \jm{CFTs}, defined on a circle of
 circumference $\ell$, regardless of what boundary conditions are chosen, as $\ket{\cB}$ approaches the boundary state \jm{$\ket{B}$}, the LREE always contains a \DD{cut-off dependent divergent} term proportional to the central charge ($c$),\begin{align}
S_{\text{div}} = \frac{\pi c \, \ell}{24 \epsilon}.  
\end{align} 
Also, if we work  {directly with the Ishibashi states}, then there is a finite piece which can be written precisely in terms of the \textit{quantum dimensions}, $\qdim_a$, as, 
\begin{align}\label{ishi}
S_{\kket{h_a}} &=\frac{\pi c \, \ell}{24 \epsilon} +  \log  \left(\frac{\qdim_a}{\cD} \right),
\end{align}
where, $\cD =( \sum_a \qdim_a^2)^{1/2}$, is the total quantum dimension. This expression above reproduces the answer for the spatial entanglement entropy corresponding to a single anyonic excitation of type $a$ in a (2+1)-$d$   in topological quantum field theory (TQFT) which has a chiral CFT describing its edge modes. The first term in \eqref{ishi} being the \textit{area law term} and the second term the \textit{topological entanglement entropy} \cite{wen2006,kitaev-preskill}. As we elaborate below, the reason why this happens is because the entanglement/modular Hamiltonian of a bulk (2+1)-$d$ theory is equal to the Hamiltonian describing a chiral CFT on its edge \cite{haldane-li,ludwig-qi,chandran, rex-prb, Cano,swingle}. This result, therefore, follows as a consequence of the correspondence relating 2$d$ CFTs and 3$d$ TQFTs \cite{Witten,Moore-Read}. 

The {Ishibashi} states $\kket{h_a}$, however, do not satisfy modular invariance when the CFT is on a finite cylinder. Imposing this constraint gives us a specific linear combination of $\kket{h_a}$, which are known as the \textit{Cardy states}, $\ket{\widetilde{l}}$ \cite{cardy, cardy-bcft}.  We derive expressions for LREE for these  states in diagonal CFTs in terms of the modular $\cS$-matrix of the CFT
\begin{align}\label{0-bc}
S_{\ket{\widetilde{h}_{l}}} &=\frac{\pi c \, \ell}{24 \epsilon}-  \sum_{j} \cS_{lj}^2 \log \left( \frac{\cS_{lj}^2}{\cS_{1j}} \right).
\end{align}
This formula clearly shows the explicit dependence of this quantity on the choice of $\ket{\cB}$ which is parametrized here by the index $l$. For the choice $l=1$, the finite piece becomes $-  \sum _a  \left( {\qdim_a^2}/{\cD^2} \right) \log  \left( {\qdim_a}/{\cD} \right)$. Once again, using the TQFT/CFT correspondence, we interpret this as the  {average topological entanglement entropy}.  We explicitly evaluate the LREE corresponding to the Cardy states for a few cases --  two minimal models describing statistical systems at criticality (the Ising model and the tricritical Ising model) and the $\widehat{su}(2)_k$ WZW model. 

\section{Left-right entanglement entropy formulation for general CFT{\tiny s} }
 We first develop a formalism to calculate LREE for an arbitrary boundary state, $\ket{ B } =\sum _{a} \psi_B^a\kket{h_a}$.  {The coefficients $\psi_B^{h_a}$ characterize the boundary condition.} $\kket{h_a}$ are the Ishibashi states,  which are solutions to the conformal boundary condition $L_n \ket{b} = \overline{L}_{-n}\ket{b}$ \cite{ishibashi}. They can be expressed in terms of the orthonormal basis $\ket{h_a, N ; k}$ and $\ket{\overline{ h_a,   N ;  k}}$. 
\begin{align}
\kket{h_a} = \sum_{N=0}^\infty  \sum_{k=1}^{d^{h_a}_N} \ket{h_a, N ; k} \otimes {\ket{\overline{ h_a,   N ;  k}}}.
\end{align}
 Here, $a$ is a label for the $a^{\rm th}$ primary  {of weight $h_a$} and the sum $N$ is over descendants. \DD{As mentioned earlier, since $\ket{B}$ is non-normalizable, we consider the regularized state $\ket{\cB} = ( e^{-\ep H}/{{(\cZ_{B} )}^{1/2}}  )\ket{B}$. The reduced density matrix associated with this state for the left-moving sector is then \cite{zayas, Miyaji:2014mca}  }
$$\rho^{(B)}_L ={\tr_R \left(	e^{-\epsilon H} \ket{B} \bra{B} e^{-\epsilon H}	\right)\over \cZ_{B}  },$$ 
where,
$
\editx{\cZ_{B}  
= \sum_{j} ({|\psi^{h_j}_B|^2})\, \chi_{{h_j}}(e^{-8\pi \frac{\epsilon}{\ell}}),}
$ is the normalization factor. The Hamiltonian here is,
$
H = \frac{2\pi}{\ell} ( L_0 + \bar{L}_0 - \frac{c}{12} ).
$
More explicitly,  {after tracing out the right-moving sector},
\editx{\begin{equation}\label{five}
\rho_L^{(B)} =  \frac{ \sum\limits_{a, N, k} |\psi_B^{h_a}|^2  e^{-8\pi \frac{\epsilon}{\ell}( h_a+ N -\frac{c}{24})}\ket{h_a, N, k }\bra{h_a, N, k} }{\cZ_{B}}.
\end{equation}}
As a check of consistency, it can be shown that the reduced density matrix becomes $\ket{0}\bra{0}$ in the $\ep \to \infty$ limit. As expected, this is the projector onto the left-moving sector's vacuum, since the ground state of the CFT on the circle is just $\ket{0}\otimes \ket{\bar{0}}$ and hence the corresponding LREE is zero.

We shall be interested in the thermodynamic limit, $\ell/\epsilon \gg 1$. Although one can directly proceed to calculate the entanglement entropy using the reduced density matrix \eqref{five}, we use the replica trick for the sake of computational simplicity. (\ref{five}) is a diagonal density matrix and can therefore be easily raised to its $n^{\text{th}}$ power. 
We require the following trace,  
\editx{
\begin{align}\label{trace-of-rho-n}
\tr_L [\rho_L^{(B)}]^n= \frac{1}{(\cZ_{B})^n}  \sum_{a,a'} |\psi_{B}^{h_a}|^{2n}  \cS_{aa'}\chi _{\null_{h_{a'}}} (e^{-\frac{  \pi \ell}{2n\epsilon}}). 
\end{align}
}
Here, $\chi_{h_a}$ are the characters  of the highest weight representations corresponding to the primaries ${h_a}$. For unitary CFTs, they admit the expansion,  
$ 
\chi_h ( q) \ = \ \tr_{\mathcal{V}_h}\left( q^{L_0-\frac{c}{24}} \right)  \ = \   q^{\,-\frac{c}{24}} \sum_{N \geq 0} d^h_N \, q^{\, h+N}
$ \cite{yellowbook}. Here, $d^h_N$ counts the degeneracy of descendants at each level $N$.
In equation \eqref{trace-of-rho-n} above, we have also used the modular transformation, $\chi_{h_a}(e^{-8 \pi n\frac{\epsilon}{\ell}})=\sum_{a'} \cS_{aa'}\chi _{h_{a'}} (e^{-\frac{  \pi \ell}{2n\epsilon}})$, to facilitate a small $\ep/\ell$ expansion. 


%
%

Using the  expansion for the characters in \eqref{trace-of-rho-n}, we obtain the following trace
\begin{align}
\tr_L  [\rho_L^{(B)} ]^n=& \ \exp\left[  {\frac{\pi \ell}{2\epsilon} \left(\frac{1}{n}-n\right)\frac{c}{24}} \right] \  \frac{\mathcal{F}(n,B)}{[\mathcal{F}(1,B)]^n} ,  
\end{align}
where,  
\begin{align}
\mathcal{F}(n,B) \ \stackrel{\ep/\ell \to 0}{=} \  \ {   \sum\limits_{a,a'}  {|\psi_B^{h_a}|^{2n}}      \cS_{aa'} \,  \sum\limits_{N \geq 0} d^{h_{a'}}_N \tq^{(h_{a'}+N)/n}    } \nn ,
\end{align}
with $\tq=\exp\left(-\frac{\pi \ell }{2\epsilon}\right)$.

%
By taking the derivative w.r.t.~$n$, we explicitly obtain the entanglement entropy to be  
\begin{align}
\label{uni-formula}
   & S_{\ket{B}} = -\, \pd_n \tr [\rho_L^{(B)}]^n \Big|_{n \to 1} , \\
=& \ \ \frac{\pi c\,\ell}{24 \,\epsilon}   - \frac{\sum\limits_a \cS_{1a} |\psi_{B}^{h_a}|^2 \log(|\psi_{B}^{h_a}|^2 )}{ \sum\limits_a \cS_{1a  }|\psi_{B}^{h_a}|^2 } + \log  \sum\limits_a \cS_{1a} |\psi_{B}^{h_a}|^2 . \nn 
\end{align}
This clearly shows the presence of the divergent piece in the left-right entanglement entropy for any choice of $\psi_B^{h_a}$. Notice choosing $\ket{B}$ to be an Ishibashi state $\kket{h_j}$, gets rid of the second term and the finite piece is $\log \cS_{1j}$. 

We shall now restrict our discussion to diagonal CFTs \cite{yellowbook}. These are the theories which have the modular invariant partition functions (on the torus) to be of the form $Z=\sum_{a} |\chi_{\null_{h_a}}\hspace{-.08cm}|^2$ and have all fields in their corresponding Kac table to be present (\textit{e.g.}~minimal models of the type $(A_{p-1},A_{q-1})$ in the ADE classification). For theories of this kind, the Cardy states can be written as \cite{cardy-bcft,gaberdiel-lectures2,about-S}, 
\begin{align}\label{diag-bdy-state}
\ket{B} \equiv \ket{{\widetilde{h}_l}} = \sum_j \frac{\cS_{lj}}{\sqrt{\cS_{1j}}} \kket{h_j},
\end{align}
i.e.~$\psi_{l}^j = \cS_{lj}/\sqrt{ \cS_{1j} }$. The result \eqref{uni-formula} can now be expressed fully in terms of the modular $\cS$-matrix. Using the real, symmetric and unitary properties of $\cS$ it is easy to see that   only the second term gives a finite contribution. Thus, for generic Cardy states in diagonal CFTs, the result for the LREE  is,
\begin{align}\label{general-lree}
S_{\ket{\widetilde{h}_l}} =\frac{\pi c \ell}{24 \epsilon}-   \sum_{j} \cS_{lj}^2 \log \left( \frac{\cS_{lj}^2}{\cS_{1j}} \right).
\end{align}
It is worthwhile noting here that the finite contribution in \eqref{uni-formula}, \eqref{general-lree} is completely independent of the scales involved in the system.

\section{Quantum dimensions and  topological entanglement entropy}

\def\topo{{\text{topo}}}
Quite interestingly, if we specifically choose the boundary state $\ket{B}$ to be an Ishibashi state, $\kket{h_j}$ 
-- equation \eqref{uni-formula}  with $\psi_{B}^{h_a}=\delta_{aj}$ -- then the LREE  can be completely recast in terms of the \textit{quantum dimensions} of the CFT. The quantum dimensions $\lbrace \qdim_j \rbrace$ and the total quantum dimension $\cD$ are defined in terms of the elements of the modular $\cS$ matrix as follows \cite{moore-naturality,Verlinde-fusion,topoEE} 
\begin{align}
\qdim_j = \frac{\cS_{1j}}{\cS_{11}} \quad , \quad \cD = \sqrt{\sum_j \qdim_j^2} = \frac{1}{\cS_{11}} \ ,
\end{align}
(the second relation follows from unitarity of $\cS$).
Using the above definition, 
the left-right entanglement entropy corresponding to $\kket{h_j}$ is  \cite{q-dim}

\begin{align}\label{0-universal}
S_{\kket{h_j}} =\frac{\pi c \ell}{24 \epsilon} + \log  \frac{\qdim_j}{\cD}  
\end{align}

It is well-known that gapless edge modes of (2+1)-$d$ TQFT can be described by a chiral (1+1)-$d$ rational conformal field theory  \cite{Witten,Moore-Read}. Hence, we interpret equation \eqref{0-universal} \editx{in terms of the} bulk quantities as follows. The first term in \eqref{0-universal} has the same structure of the area law of real-space entanglement entropy $\sS_\area (= \alpha \, \ell/\epsilon)$ in two spatial dimensions.  $\ep$, therefore, corresponds to  the spacing of the lattice on which the TQFT lives. The non-universal constant factor is $\pi c /24$. 
 Moreover, the second term, $  \log ({\qdim_j}/{\cD})$, is exactly the \textit{topological entanglement entropy},  $\sS_\topo(j)$, of anyons of type $j$  \cite{kitaev-preskill,wen2006,dong}. This term is universal and captures global features of entanglement in (2+1)-$d$ TQFTs. 
We can therefore rewrite equation \eqref{0-universal} as
\begin{align} \label{main}
S_{\kket{{h}_j}} \ = \  \sS_{\text{area}} + \sS_{\text{topo}}(j).
\end{align}
 Therefore, \textit{the left-right entanglement entropy for Ishibashi states precisely reproduces the spatial-entanglement entropy in (2+1)-$d$}  \cite{kitaev-preskill,wen2006}. 

\begin{figure}[t] 
\includegraphics[scale=.40]{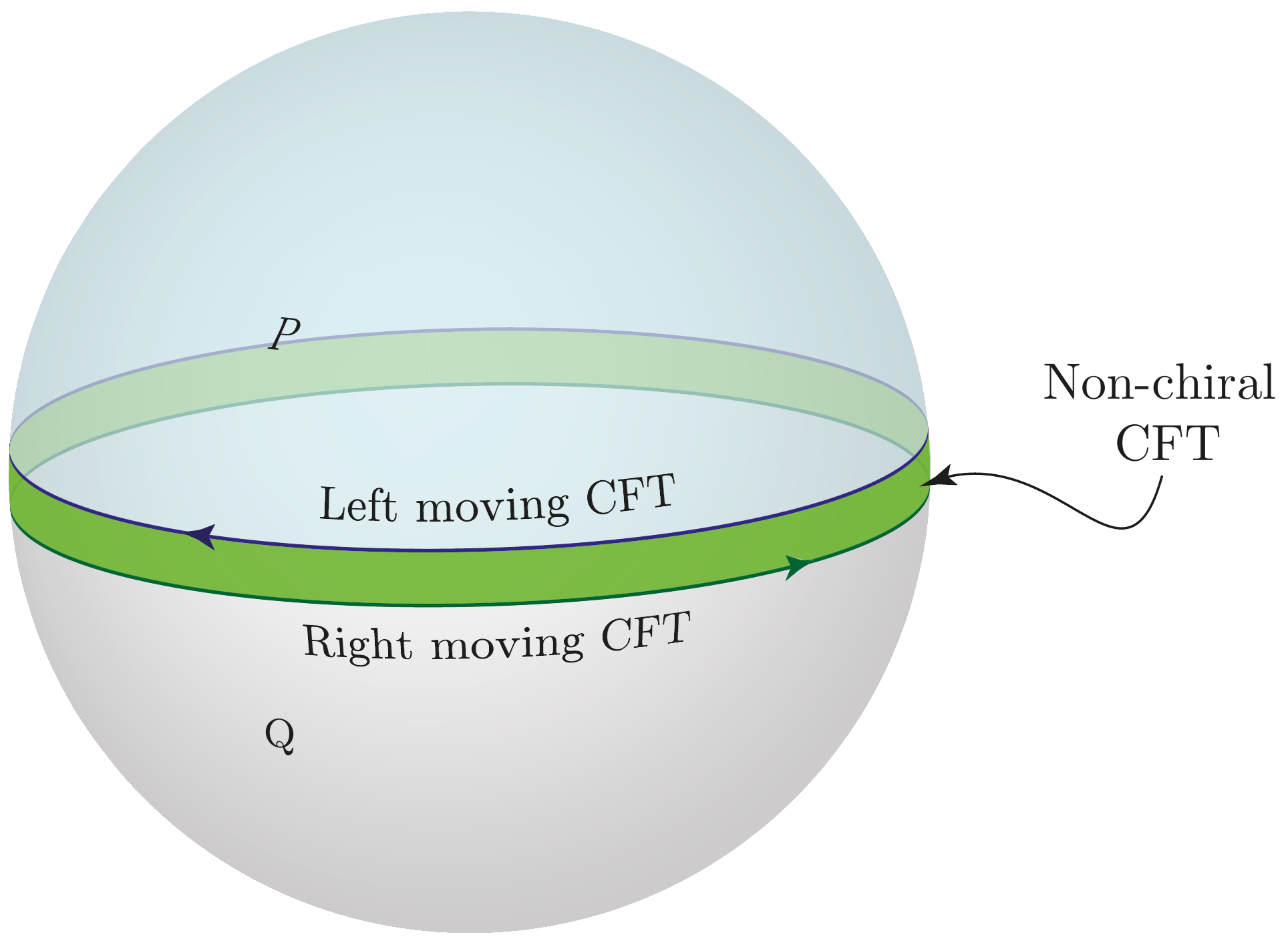}
\caption{A TQFT on a sphere which is spatially bipartitioned into the upper and lower hemispheres. A non-chiral CFT lives in the interedge region. Calculating the left-right entanglement entropy of this CFT is equivalent to calculating the real-space entanglement entropy in (2+1)-$d$.}
\end{figure}
 
The  result \eqref{main} can be understood from the equivalence of the modular Hamiltonian of the  bulk theory with the Hamiltonian of the chiral CFT living on its edge \cite{haldane-li,ludwig-qi,chandran, rex-prb, Cano,swingle}. Let us review this equivalence briefly. Consider anyons in the topological sector $j$ living on a sphere $S^2$ (see Fig.~1) -- of radius $\ell/(2\pi)$ --  which is bipartitioned into its upper ($P$) and lower ($Q$) hemispheres. The edge of $P$ ($Q$) has a chiral (anti-chiral) CFT. 
The full Hamiltonian of this system is, $H=H_P +H_Q + H_{PQ}$. Here, $H_{PQ}$ is the RG-relevant  interedge coupling between the left and right modes. 
It can then be shown, entanglement properties between $P$ and $Q$ can be mapped to those between the left- and right-moving edge modes of the non-chiral CFT living in the interedge region  \cite{ludwig-qi, rex-prb, Cano}. Therefore, tracing out the subsystem $Q$ is tantamount to tracing out the right-moving modes. 

The result for the LREE, for the specific Cardy state $\ket{\widetilde{0}}$ (equation \eqref{diag-bdy-state} with $l=1$), also admits an expression in terms of the quantum dimensions. From \eqref{main},
\begin{align} \label{avg-formula}
S_{\ket{\widetilde{0}}} =\frac{\pi c \ell}{24 \epsilon} - \sum_j  \left({{\qdim_j}\over{\cD}}\right)^2   \log { {\qdim_j}\over{\cD}  }
\end{align}
The linear combination of Ishibashi states in $\ket{\widetilde{0}}$ is such that the probability of $\kket{h_a}$, in the boundary state is $\qdim_a^2/\cD^2$ \cite{average}. Equivalently, from TQFT fusion rules, it follows that the factor $\qdim_j^2/\cD^2$ is also the probability of finding a quasiparticle of type $j$ in an anyonic gas at steady state \cite{preskill,wangTQC}. The second term can then be interpreted as the average topological entanglement entropy.

\section{Examples}
As is evident from the formula \eqref{general-lree}, the LREE can be readily calculated for the Cardy states \eqref{diag-bdy-state} if we know the central charge and the modular $\cS$ matrix. \jm{The calculation of LREE for the free boson theory \cite{zayas} indeed contains the identical form of the leading divergent term as predicted. }
In this section, we explicitly evaluate the LREE for the Ising model, the triticritical Ising model and the $\widehat{su}(2)_k$ WZW model. 
\\


\subsection{Ising model}

The Ising model, which  corresponds to the Virasoro minimal model $\cM(4,3)$, has a central charge of $c=1/2$. It has three operators : the identity $\mathbbm{I}$, the Ising spin $\sigma$ and the thermal operator $\varepsilon$. 
These primaries have weights $
0 , \ 1/2 , \ 1/16
$ respectively. \DD{This model is also equivalent to a free massless Majorana fermion.} 
The modular $\cS$ matrix can be found in \cite{yellowbook,isingdual}. 

Following are the results for the left-right entanglement entropy of the Ising model. \\
\renewcommand{\arraystretch}{1.5}
\begin{center}
\begin{tabular}{  c l } 
\hline
\ \ \ \ Boundary condition\ \ \ \ & \ \ \ \ \ \ \ \ \ \ LREE \ \   \\ 
\hline
\hline
$\ket{\widetilde 0}$ & $ (\pi \ell)/(48 \ep) +\frac{3}{4} \log 2 \ \ \ \  $ \\
$\ket{\widetilde{ \frac{1}{2}}}$ & $ (\pi \ell)/(48 \ep)+\frac{3}{4} \log 2 \ $ \\
$\ket{\widetilde{ \frac{1}{16}}}$ & $(\pi \ell)/(48 \ep) $ \\
\hline
\end{tabular}
\end{center} \null 
\DD{The chiral Ising model describes edge excitations of the $p+ip$ superconductor \cite{topoEE}. Thus, from equation \eqref{avg-formula}, the average topological entanglement entropy,  of the bulk quasiparticles in the $p+ip$ superconductor can be read out as  
\begin{align}
\langle S_{\text{topo}} \rangle_{(p+ip) \rm SC}= -  \frac{3}{4} \log 2.
\end{align}  
}
\subsection{Tricritical Ising model}
Our second example is the tricritical Ising model. This is the Virasoro minimal model $\cM(5,4)$ with $c=7/10$. It has six primary operators -- an identity ($\mathbbm{I}$), three thermal operators ($\varepsilon, \ \varepsilon ' , \ \varepsilon ''$)  and two Ising spins ($\sigma, \ \sigma '$)  -- with conformal dimensions
$
0  , \  {1}/{10}   , \ {3 / 5}   , \ {3/ 2}   , \ {3 / 80}   
$ and $ {7 / 16} $.

The results for the LREE, for various boundary states, are listed in the following table. (Here, $s_1=\frac{1}{\sqrt{5}}\sin\frac{2\pi}{5}$ and $s_2=\frac{1}{\sqrt{5}}\sin\frac{4\pi}{5}$. These appear as entries in the modular $\cS$ matrix \cite{yellowbook,isingdual}).
\begin{widetext}
\renewcommand{\arraystretch}{1.5} 
\begin{center}
\begin{tabular}{  c l } 
\hline
 Boundary condition\ \ & \ \ \ \ \ \ \ \ \ \ \ \ \ \ \ \ \ \ \ \ \ \ \ \ \ \ \ \ \ \ \ \ \ \ \ \ \ \ \ \ LREE \ \   \\ 
\hline
\hline
$\ket{\widetilde{0}}$, $\ket{\widetilde{\frac{3}{2}}}$& $ (7\pi \ell)/(240\ep) -4({\mathit{s}_1^2 \left(\log \mathit{s}_1 \right)+\mathit{s}_2^2 \left(\log\mathit{s}_2  \right)})-\frac{1 }{4}\log 2  $ \\
${\ket{\widetilde{\frac{1}{10}}}}, \ {\ket{\widetilde{\frac{3}{5}}}}$ & $  (7\pi \ell)/(240\ep)  +4({\left(\mathit{s}_2^2-2 \mathit{s}_1^2\right) \left( \log \mathit{s}_1\right)+\left(\mathit{s}_1^2-2 \mathit{s}_2^2\right) \left(\log\mathit{s}_2  \right)})-\frac{1 }{4}\log 2 \ \ \ $ \\
$\ket{\widetilde{\frac{7}{16}}}$ & $  (7\pi \ell)/(240\ep)  -4({\mathit{s}_1^2 \left(\log \mathit{s}_1 \right)+\mathit{s}_2^2 \left(\log\mathit{s}_2  \right)})-\log 2  $ \\
${\ket{\widetilde{\frac{3}{80}}}}$ & $  (7\pi \ell)/(240\ep) +4({\left(\mathit{s}_2^2-2 \mathit{s}_1^2\right) \left( \log \mathit{s}_1\right)+\left(\mathit{s}_1^2-2 \mathit{s}_2^2\right) \left(\log\mathit{s}_2  \right)})- \log 2  $ \\
\hline
\end{tabular}
\end{center}
\end{widetext}
\def\suck{\widehat{su}(2)_k}
\subsection{The $\widehat{su}(2)_k$ WZW model}

The $\suck$ WZW model has a  central charge of $c=k/(k+2)$. From the knowledge of the modular $\cS$ matrix of this theory (see for e.g.~\cite{topoEE}), it is straightforward to calculate the LREE of this theory from equation \eqref{general-lree}. (We consider the diagonal theory i.e.~with a mass matrix $\mathcal{M}=\mathbbm{I}$).
We obtain
\begin{align}
 S_{\ket{\widetilde{h}_l}} =& \dfrac{\pi k \ell}{8(k+2)\ep}  \\&  - \sum _{j=1}^{k+1} \tfrac{2}{k+2} \sin ^2\left(\tfrac{\pi j l}{k+2}\right) \log \left[ {\footnotesize{   \frac{\sqrt{\frac{2}{k+2}} \sin ^2\left(\frac{\pi j l}{k+2}\right)}{\sin \left(\frac{ \pi  j}{k+2}\right)} }}\right]. \nn 
\end{align}
In a similar manner, the LREE can be calculated for WZW models of other ranks and also for coset models.

 \section{Conclusions} 
 In this work, we have derived a universal formula for the left-right entanglement entropy for a \DD{state approaching the boundary state for  }\jm{generic CFTs on a circle}. The universal formula for LREE contains \DD{a divergent piece proportional to the central charge of the CFT for any choice of boundary states}. Moreover, for the case of diagonal CFTs with a specific choice of a boundary state, we see that the finite contribution to the LREE can be fully formulated in terms of the  quantum dimensions of the theory. 
 
Our analysis also shows that this measure of entanglement precisely captures the spatial entanglement of a higher dimensional topological field theory.   This therefore serves as an alternative derivation of topological entanglement entropy and explicitly illustrates the TQFT/CFT (bulk/edge) correspondence.  
 It is worth seeking the usefulness and  implications of this connection  for the fractional quantum hall effect \cite{zhang-fqhe}, anyon superconductivity \cite{laughlin-sc,chen1989anyon,fetter1989random} and topological quantum computing \cite{KitaevTQC,preskill}. 
 
 It shall be interesting to reproduce the LREE  using the methods of AdS/CFT \cite{TakayanagiBCFT,Ryu:2006bv,verlinde}. The topological nature of 3$d$ gravity  may be useful in this regard.
 It is worthwhile mentioning here, that the topological entanglement entropy corresponding to 3$d$ black holes having $SL(2,\mathbbm{R})$ holonomies in the hyperbolic conjugacy class  (which are the analogs to quasiparticles of a specific type) have already been evaluated in \cite{verlinde}. This analysis can be hopefully extended to other conjugacy classes (parabolic and elliptic) and it may be possible to calculate the average topological entanglement entropy. Concurrently, it would be useful to evaluate the equivalent quantity in a CFT (like Liouville theory) with large central charge and check the holographic consistency.  
 %
  
 Another intriguing direction to pursue is to investigate this quantity for D$p$-branes. The methods developed in this paper and that of \cite{zayas} can be suitably utilized to calculate the LREE. We can then try to understand whether this measure of entanglement can shed any new light on string theory \cite{tt-string}. 
\begin{acknowledgements}
It is a pleasure to thank John Cardy, Sumit Das, Justin David,  \editx{Nabil Iqbal}, John McGreevy, Apoorva Patel, Alfred Shapere, Ritam Sinha and Tadashi Takayanagi  for fruitful discussions and comments on our manuscript. The work of D.D.~is supported by funds provided by the U.S. Department of Energy (D.O.E.) under cooperative research agreement DE-FG0205ER41360. The work of S.D.~is supported by a research associateship from the Indian Institute of Science. 
\end{acknowledgements}


%

\begin{thebibliography}{50}
\bibitem [{abo(0000)}]{about-S}%
  \BibitemOpen
  \href@noop {} {\bibfield  {journal} {\bibinfo  {journal} {{$\cS$ is unitary
  and hermitian in general. However, for a wide class of examples $\cS$ is real
  and symmetric. Our result \eqref{general-lree} can be straightforwardly
  generalized to the most general case by the replacement $(S_{ij})^2 \to
  |S_{ij}|^2$. These results are also true for non-diagonal minimal models
  \cite{cardy}.}}\ } (\bibinfo {year} {0000})}\BibitemShut {NoStop}%
\bibitem [{ave(0000)}]{average}%
  \BibitemOpen
  \href@noop {} {\bibfield  {journal} {\bibinfo  {journal} {{The probability of
  occurence of the same Ishibashi state at both the boundaries ($\pd P$ and
  $\pd Q$) is given by $\big|(\bra{h_i}_{\pd Q}\otimes\bra{h_i}_{\pd
  P})(\ket{\widetilde{0}}_{\pd P}\otimes\ket{\widetilde{0}})_{\pd
  Q}\big|^2=\qdim_i^2 /\cD^2$.}}\ } (\bibinfo {year} {0000})}\BibitemShut
  {NoStop}%
\end{thebibliography}



%

\end{document}